\documentclass[12pt]{article}
\usepackage{graphicx}
\usepackage{amsmath}
\newcommand{\sm}{\hbox{$\bigcirc$\kern-0.72em\hbox{\bf s} }}
\newcommand{\Id}{\hbox{\sl 1\kern-0.25em\hbox{I}}}
\newcommand{\rcorr}{\hbox{\kern-1.2em$\longrightarrow$}}
\newcommand{\lrcorr}{\hbox{\kern-1.2em$\longleftrightarrow$}}

\newcommand{\nRightarrow}{\Rightarrow\kern-1.2em\hbox{/}\kern.8em} %
\newcommand{\BB}{\hbox{I\kern-.2em\hbox{B}}} %BB
\newcommand{\DD}{\hbox{I\kern-.2em\hbox{D}}} %DD
\newcommand{\FF}{\hbox{I\kern-.2em\hbox{F}}} %FF
\newcommand{\NN}{\hbox{I\kern-.2em\hbox{N}}}  %Naturali
\newcommand{\ZZ}{{{\rm Z}\kern-.28em{\rm Z}}} %Interi
\newcommand{\RR}{\mathop{{\rm I}\kern-.2em{\rm R}}\nolimits} %Reali
\newcommand{\RRe}{\mathop{{\rm I}\kern-.2em{\rm Re}}\nolimits} %Reali
\newcommand{\QQ}{\hbox{l\kern-.36em\hbox{Q}}}  %Razionali
\newcommand{\CC}{\hbox{{\textsf I}\kern-.47em\hbox{C}}}
\newcommand{\nop}{\hbox{{\textsf I}\kern-.47em\hbox{O}}}

\def\TREV{{{}^\triangleleft\kern-1.5pt\texttt{T}}}
\def\trev{{{}^\triangleleft\kern-3.2pt\texttt{t}}}
\def\SREV{{{}_\triangleleft\kern-2pt\texttt{S}}}
\def\srev{{{}_\triangleleft\kern-2.2pt\texttt{s}}}

\begin{document}
\title{Group Theoretical Settling\\
of Spin Zero Relativistic Particle Theories}
\author{Giuseppe Nistic\`o
\\
{\small Dipartimento
di Matematica e Informatica, Universit\`a della Calabria, Italy}\\
{\small and}\\
{\small
INFN -- gruppo collegato di Cosenza, Italy}\\
{\small email: giuseppe.nistico@unical.it} } \maketitle
\abstract{In order to avoid the difficulties encountered by  relativistic quantum theory
of single particles, we pursue a deductive development of the theory from physical principles,
without canonical quantization, by making use of group-theoretical methods.
Our work has pointed out the necessity of new classes of irreducible
representations of the Poincar\'e group the quantum theory of a particle can be based on.
For spin 0 particle, four inequivalent theories are completely determined,
with fundamental differences with respect to Klein-Gordon theory.}
%%%%%%%%%%%%%%%%%%%%%%%%%%%%%%%%%%%%%%%%%%%%%%%%%%%%%%%%%%%%%%%%%%%%%%%%%%%%%%%%%%%%
\section{Introduction}
{\sl Canonical quantization} was the primary method for formulating {\sl specific
relativistic particle theories} \cite{Klein},\cite{Fock},\cite{Gordon};
despite the successful results in the non relativistic case, the problems encountered by
{\sl relativistic particle theories} yielded by canonical quantization ultimately led theoretical physicists to turn on quantum field theory
to model elementary particle physics \cite{WeinBook}.
\par
This state of affairs
has its roots in the methodological features of canonical quantization. In order to formulate
the quantum theory of a specific physical system, canonical quantization prescribes to start from its {\sl classical theory}. For a particle
we could start
from its Hamiltonian classical theory, where the
position's coordinates $(q_1,q_2,q_3)\equiv{\bf q}$ are dynamical variables, with
conjugate momenta ${\bf p}\equiv(p_1,p_2,p_3)$ and with Hamiltonian function $h({\bf q},{\bf p},t)$;
then the prescriptions of canonical quantization dictate to replace the dynamical variables $q_j$ and their momenta $p_j$ with
operators $Q_j$ and $P_j$, and to replace the Poisson brackets $\{\;,\;\}$ of the classical theory with operator's commutators $i[\;,\;]$.
We see that this procedure provides no deductive path that leads to results from physical principles.
For this reason the real causes of problematic or inconsistent
predictions cannot be singled out to be remedied, in general.
\par
The aim of this work is to pursue an approach {\sl alternative} to canonical quantization,
which formulates the theories
through a {\sl deductive development from physical principles}.
This methodological commitment should prevent from the problems occurring with canonical quantization.
\par
For an isolated system, we shall assume as physical principle the invariance with respect to Poincar\'e transformations,
which implies the existence of a projective representation of the Poincar\'e group in the Hilbert space of the theory.
Consequently, the first step of our approach, in sections 2 and 3, has been the identification of the class of irreducible projective representations of Poincar\'e group, that
turned out to be richer
than that considered in the literature; in particular, also representations with anti-unitary space inversion operator are identified.
\par
In section 4, coherently with our methodological commitments we identify the quantum theory of an elementary free particle,
specific example of isolated system,
by imposing the transformation properties
of the position observable, expressed by means of a suitably conceived notion of {\sl quantum transformation}.
Accordingly,
the identification is addressed by selecting
the irreducible representations that admit such a position operator.
\par
For spin zero particle we identify four inequivalent complete theories.
Though Klein-Gordon equation can be derived from the present theories, there are fundamental differences. In particular, two of the theories
are based on irreducible representations of the Poincar\'e group belonging to the new classes.
The theories do not suffer the shortcomings of Klein-Gordon theories, such as negative density of position probability.
Moreover,
the inconsistency of the four density pointed out by Barut and Malin \cite{BM68} is avoided.
\section{Notation and mathematical tools}
We shall make use of the following mathematical structures developable within the formalism of a complex and separable Hilbert space $\mathcal H$, that
are of general interest also in quantum theory.
\begin{description}
\item[\quad-]
The set $\Omega({\mathcal H})$ of all self-adjoint operators of $\mathcal H$; in a quantum theory these operators represent
{\sl quantum observables}.
\item[\quad-]
The lattice $\Pi({\mathcal H})$ of all projections operators of $\mathcal H$; in a quantum theory they represent observables with spectrum $\{0,1\}$.
\item[\quad-]
The set $\Pi_1({\mathcal H})$ of all rank one orthogonal projections of $\mathcal H$.
\item[\quad-]
The set ${\mathcal S}({\mathcal H})$ of all density operators of $\mathcal H$;
in a quantum theory these operators represent {\sl quantum states}.
\item[\quad-]
The set ${\mathcal V}({\mathcal H})$ of all unitary  or anti-unitary operators of the Hilbert space $\mathcal H$.
\item[\quad-]
The set ${\mathcal U}({\mathcal H})$ of all unitary operators of $\mathcal H$; trivially,
${\mathcal U}({\mathcal H})\subseteq{\mathcal V}({\mathcal H})$ holds.
\end{description}
The following definition introduces generalized notions of group representation.
\vskip.5pc\noindent
{\bf Definition 2.1.} {\sl Let $G$ be a separable, locally compact group with identity element $e$. A correspondence
$U:G\to{\mathcal V}({\mathcal H})$, $g\to U_g$, with $U_e=\Id$, is a generalized projective representation of $G$
if the following conditions are satisfied.
\begin{description}
\item[\;{\rm i)}\;\;]
A complex function $\varsigma:{G}\times{G}\to{\CC}$,
called multiplier,
exists such that $U_{g_1g_2}=\varsigma(g_1,g_2)U_{g_1}U_{g_2}$; the modulus $\vert\varsigma(g_1,g_2)\vert$ is always 1, of course;
\item[\;{\rm ii)}\;]
for all $\phi,\psi\in\mathcal H$, the mapping $g\to\langle U_g\phi\mid\psi\rangle$ is a Borel function in $g$.
\end{description}\noindent
Whenever $U_g$ is unitary for all $g\in G$, $U$ is called
projective representation, or $\varsigma$-representation.
\par
A generalized projective representation is said to be continuous if for any fixed $\psi\in\mathcal H$
the mapping $g\to U_g\psi$ from $G$ to $\mathcal H$ is continuous with respect to $g$.
}
\vskip.5pc\noindent
%All multipliers must satisfy the following conditions \cite{c11}.
%$$
%\sigma(e,g)=1, \quad \frac{\sigma(h_1,h_2)}{\sigma(h_2,h_3)}=\frac{\sigma(h_1,h_2h_3)}{\sigma(h_1h_2,h_3)}, \quad
%\hbox{for all }g,h_1,h_2,h_3\in{\Upsilon}.
%\eqno(31)
%$$
In \cite{N1} we have proved that the following statement holds.
\vskip.5pc\noindent
{\bf Proposition 2.1.}
{\sl If $G$ is a connected group, then every continuous generalized projective representation of $G$ is a projective
representation, i.e. $U_g\in{\mathcal U}({\mathcal H})$, for all $g\in G$.}
\vskip.5pc
Given any vector $\underline x=(x_0,{\bf x})\in\RR^4$, we call $x_0$ the {\sl time component} of $\underline x$
and ${\bf x}=(x_1,x_2,x_3)$ the {\sl spatial component} of $\underline x$.
The proper orthochronous Poincar\'e group ${\mathcal P}_+^\uparrow$ is the separable locally compact
group of all transformations of $\RR^4$
generated by the ten one-parameter sub-groups ${\mathcal T}_0$, ${\mathcal T}_j,{\mathcal R}_j$, ${\mathcal B}_j$, $j=1,2,3$,
of time translations, spatial translation, proper spatial rotations and Lorentz boosts, respectively.
The Euclidean group $\mathcal E$ is the sub-group generated by all ${\mathcal T}_j$ and ${\mathcal R}_j$.
The sub-group generated by
${\mathcal R}_j$, ${\mathcal B}_j$ is the proper orthochronous Lorentz group ${\mathcal L}_+^\uparrow$ \cite{BarBook}.
It does not include time reversal $\trev$ and space inversion $\srev$.
Time reversal $\trev$ transforms $\underline x=(x_0,{\bf x})$ into $(-x_0,{\bf x})$; space inversion $\srev$
transforms $\underline x=(x_0,{\bf x})$ into $(x_0,-{\bf x})$.
The group generated by $\{{\mathcal P}_+^\uparrow, \trev,\srev\}$ is the separable and locally compact Poincar\'e group $\mathcal P$.
By ${\mathcal L}_+$ we denote the subgroup generated by ${\mathcal L}_+^\uparrow$ and $\trev$, while ${\mathcal L}^\uparrow$
denotes the subgroup generated by ${\mathcal L}_+$ and $\srev$; analogously, ${\mathcal P}_+$ denotes
the subgroup generated by ${\mathcal P}_+^\uparrow$ and $\trev$, while ${\mathcal P}^\uparrow$ is the subgroup
generated by ${\mathcal P}_+^\uparrow$ and $\srev$.
\vskip.5pc
All sub-groups ${\mathcal T}_0$, ${\mathcal T}_j,{\mathcal R}_j$, ${\mathcal B}_j$ of ${\mathcal P}_+^\uparrow$ are additive;
in fact, ${\mathcal B}_j$ is not additive with respect to the parameter
{\sl relative velocity} $u$, but it is additive with respect to the parameter $\varphi(u)=\frac{1}{2}\ln\frac{1+u}{1-u}$.
Then, according to Stone's theorem \cite{c9}, for every continuous projective representation $U$ of ${\mathcal P}_+^\uparrow$ there exist
ten self-adjoint generators $P_0$, $P_j$, $J_j$, $K_j$, $j=1,2,3$, of the ten
one-parameter unitary subgroups $\{e^{iP_0t}\}$,   $\{e^{-i{P}_j a_j},\,a\in{\RR}\}$,
$\{e^{-i{J}_j \theta_j},\,\theta_j\in{\RR}\}$, $\{e^{-i{K}_j \varphi(u_j)},\,u_j\in{\RR}\}$
that represent the one-parameter
sub-groups ${\mathcal T}_0$, ${\mathcal T}_j,{\mathcal R}_j$, ${\mathcal B}_j$ according to the projective representation
$g\to {U}_g$ of the Poincar\'e group ${\mathcal P}_+^\uparrow$.
\subsection{Characterizations of irreducible representations of $\mathcal P$}
Now we state properties of {\sl irreducible} generalized projective representation $U$ of ${\mathcal P}$
whose restriction to ${\mathcal P}_+^\uparrow$ is continuous, so that $U_g\in{\mathcal U}({\mathcal H})$
for all $g\in{\mathcal P}_+^\uparrow$, according to Prop. 2.1.
The proofs of statements not given in the present work can be found in \cite{PROOFS}.
\vskip.5pc
While $U_g\in{\mathcal U}({\mathcal H})$ if $g\in{\mathcal P}_+^\uparrow$, since $\trev$ and $\srev$ are not connected to ${\mathcal P}_+^\uparrow$,
the time reversal operator$\TREV=U_\trev$ and and the space inversion operator $\SREV=U_\srev$ can be unitary or anti-unitary; now we see how this character is related
to spectral properties of $P_0$.
\vskip.5pc\noindent
{\bf Proposition 2.2.}
{\sl
If a generalized projective representation $U:{\mathcal P}\to{\mathcal U}({\mathcal H})$
is {\sl irreducible}, then a real numbers $\eta$ exists such that $P_0^2-{\bf P}^2=\eta\Id$.
This statement holds for an irreducible projective representation
$U:{\mathcal P}_+^\uparrow\to{\mathcal U}({\mathcal H})$ too.}
\vskip.5pc\noindent
In the present work we consider only {\sl irreducible} generalized projective representations of ${\mathcal P}$ with {\sl positive} parameter $\eta>0$.
\par
The following proposition establishes that
the spectrum $\sigma(P_0)$ of $P_0$
must be one of three definite subsets $I^+_\mu$, $I^-_\mu$, $I^+_\mu\cup I^-_\mu$ of $\RR$, where
$\mu$ denotes the {\sl positive} square root $\sqrt{\eta}$, and $I^{+}_{\mu}=[\mu,\infty)$, $I^{-}_{\mu}=(-\infty,-\mu]$.
The different possibilities are related to the unitary or anti-unitary character of $\SREV$ and $\TREV$.
\vskip.5pc\noindent
{\bf Proposition 2.3.}
{\sl
If $U:{\mathcal P}\to {\mathcal V}({\mathcal H})$ is an irreducible generalized projective representation,
then there are only the following mutually exclusive possibilities for the spectrum $\sigma(P_0)$ of $P_0$.
\vskip.3pc\noindent
({\bf u})\quad $\sigma(P_0)=I^{+}_{\mu}$ and $\sigma(P_0)=[\mu,\infty)$, {\it up} spectrum;
\vskip.3pc\noindent
({\bf d})\quad $\sigma(P_0)=I^{-}_{\mu}$ and $\sigma(P_0)=(-\infty,-\mu]$, {\it down} spectrum;
\vskip.3pc\noindent
({\bf s})\quad $\sigma(P_0)=I^{+}_{\mu}\cup I^{-}_{\mu}$  and $\sigma(P_0)=[\mu,\infty)\cup(-\infty,-\mu]$, {\it symmetrical} spectrum.
\vskip.4pc\noindent
If {$\TREV$} is anti-unitary and {\rm $\SREV$} is unitary, then either
$\sigma(P_0)=I_\mu^+$ or $\sigma(P_0)=I_\mu^-$, and hence $\sigma(P_0)=I_\mu^+\cup I_\mu^-$
cannot occur.
\par\noindent
If $\TREV$ is unitary then $\sigma(P_0)=
I_\mu^+\cup I_\mu^-$, independently of $\SREV$.
\par\noindent
If $\SREV$ is anti unitary then $\sigma(P_0)=I_\mu^+\cup I_\mu^-$, independently of $\TREV$.}
\vskip.5pc
Given an irreducible generalized projective representation $U$ of $\mathcal P$, we define the projection operators
$E^{\pm}=\int_{I_\mu^\pm}p_0dE_{p_0}^{(0)}$ where $E_{p_0}^{(0)}$ is the resolution of the identity of $P_0$.
In \cite{PROOFS} we prove the following proposition.
\vskip.5pc\noindent
{\bf Proposition 2.4.}
{\sl
In an irreducible generalized projective representation $U:{\mathcal P}\to{\mathcal V}({\mathcal H})$
the relation $[E^\pm,U_g]=\nop$ holds for all $g\in{\mathcal P}_+^\uparrow$ .}
\vskip.5pc\noindent
According to Prop. 2.4, in the case of symmetrical spectrum $\sigma(P_0)=I_\mu^+\cup I_\mu^-$,
the restriction $U\mid_{{\mathcal P}_+^\uparrow}$ is always reduced by $E^+$ into
$U^+\mid_{{\mathcal P}_+^\uparrow}=E^+U\mid_{{\mathcal P}_+^\uparrow}E^+$
\;and \;$U^-\mid_{{\mathcal P}_+^\uparrow}=E^-U\mid_{{\mathcal P}_+^\uparrow}E^-$.
\par\noindent
If $\sigma(P_0)=I_\mu^+$ (resp., $\sigma(P_0)=I_\mu^-$),
then $U\mid_{{\mathcal P}_+^\uparrow}=U^+\mid_{{\mathcal P}_+^\uparrow}$ (resp., $U\mid_{{\mathcal P}_+^\uparrow}=U^-\mid_{{\mathcal P}_+^\uparrow}$).
\par
In any case, the reduction $U^\pm\mid_{{\mathcal P}_+^\uparrow}$ can be reducible or not.
\section{The irreducible representations $U$ of $\mathcal P$ with $U^+\mid_{{\mathcal P}_+^\uparrow}$ irreducible}
In general \cite{PROOFS}, every irreducible projective representation of $\mathcal P$ with $\eta>0$ is characterized by a parameter $s\in\frac{1}{2}\NN$, called
{\sl spin} parameter, and the number $\mu=\sqrt{\eta}>0$.
\par
Props. 2.3 and 2.4 imply that the irreducible representations $U$ of $\mathcal P$ can be classified according to the spectrum $\sigma(P_0)$ and to
the reducibility/irreducibility of $U^\pm\mid_{{\mathcal P}_+^\uparrow}$.
In the present section we present the complete identification of the possible irreducible generalized projective representations $U$ of $\mathcal P$ with
$U^\pm\mid_{{\mathcal P}_+^\uparrow}$ {\sl irreducible}, as determined in \cite{PROOFS}.
\subsection{The case $\sigma(P_0)=I_\mu^\pm$ with $U^\pm\mid_{{\mathcal P}_+^\uparrow}$ irreducible}
Once fixed $s$, $\mu$, if $\sigma(P_0)=I_\mu^\pm$,
{\sl modulo unitary isomorphisms} there is only one irreducible projective representation
of ${\mathcal P}_+^\uparrow$ with $\sigma(P_0)=I_\mu^+$ and only one with $\sigma(P_0)=I_\mu^-$, that we briefly present.
The Hilbert space of the projective representation is the space $L_2(\RR^3,\CC^{2s+1},d\nu)$ of the functions
$\psi:\RR^3\to \CC^{2s+1}$,
${\bf p}\to\psi({\bf p})$, square integrable with respect to the measure
$d\nu({\bf p})=\frac{dp_1dp_2dp_3}{\sqrt{\mu^2+{\bf p}^2}}$.
\vskip.8pc\noindent
For the case $\sigma (P_0)=I^+_\mu$, the following statements hold.
\begin{itemize}
\item[--]
The generators $P_j$ are the multiplication operators defined by
$(P_j\psi)({\bf p})=p_j\psi({\bf p})$;
\item[--]
$(P_0\psi)({\bf p})=p_0\psi({\bf p})$ where $p_0=+\sqrt{\mu^2+{\bf p}^2}$,
($P_0$ has a positive spectrum);
\item[--]
the generators $J_k$ are given by $J_k=i\left(p_l\frac{\partial}{\partial p_j}-p_j\frac{\partial}{\partial p_l}\right)+S_k$,
$(k,l,j)$ being a cyclic permutation of $(1,2,3)$,
where $S_1,S_2,S_3$ are the self-adjoint generators of an irreducible projective representation $L:SO(3)\to\CC^{2s+1}$
such that $S_1^2+S_2^2+S_3^2=s(s+1)\Id$;
hence, they can be fixed to be the three spin operators of $\CC^{2s+1}$;
\item[--]
the generators $K_j$ are given by
$K_j=ip_0\frac{\partial}{\partial p_j}-\frac{({\bf S}\land {\bf p})_j}{\mu+p_0}$;
\item[--]
the unitary space inversion operator and the anti-unitary time reversal operator are
$$
\SREV=\Upsilon,\quad\hbox{and}\quad\TREV=\tau{\mathcal K}\Upsilon,
\eqno(1)
$$
\item[]
where
\item[] - $\Upsilon$ is the unitary operator defined by
$(\Upsilon\psi)({\bf p})=\psi(-{\bf p})$,
\item[]
-
$\tau$ is a unitary matrix of \,$\CC^{2s+1}$ such that $\tau {\overline S_j}\tau^{-1}= -{S_j}$, for all $j$;
such a matrix always exists and it is unique up a complex factor of modulus 1;
moreover,
if $s\in\NN$ then $\tau$ is symmetric and $\tau\overline\tau=1$, while if $s\in\left(\NN+\frac{1}{2}\right)$ then $\tau$ is
anti-symmetric and $\tau\overline\tau=-1$ \cite{c13};
\item[]
-
$\mathcal K$ is the anti-unitary complex conjugation operator defined by ${\mathcal K}\psi({\bf p})=\overline{\psi({\bf p})}$.
\end{itemize}
\vskip.5pc\noindent
For the irreducible projective representation with $\sigma(P_0)=I_\mu^-$,
the generators $P_j$, $J_k$, $\TREV$ and $\SREV$ are identical to the case  $\sigma(P_0)=I_\mu^-$; $P_0$ and $K_J$ change sign.
\subsection{The case $\sigma(P_0)=I_\mu^+\cup I_\mu^-$ with $U^+\mid_{{\mathcal P}_+^\uparrow}$ irreducible}
If $\sigma(P_0)=I_\mu^+\cup I_\mu^-$, once fixed $s$ and $\mu$,
modulo unitary isomorphisms, the Hilbert space of the representation
is ${\mathcal H}=L_2(\RR^3,\CC^{2s+1},d\nu)\oplus L_2(\RR^3,\CC^{2s+1},d\nu)$,
where $E^+{\mathcal H}=L_2(\RR^3,\CC^{2s+1},d\nu)\oplus\{0\}$ and $E^-{\mathcal H}=\{0\}\oplus L_2(\RR^3,\CC^{2s+1},d\nu)$ \cite{PROOFS}.
It is convenient to represent each vector $\psi\in{\mathcal H}$, $\psi=\psi^++\psi^-$, with $\psi^+=E^+\psi$ and $\psi^-=E^-\psi$, as a column vector
$\psi=\left[\begin{array}{c}\psi^+\cr \psi^-\end{array}\right]$, where $\psi^\pm\in L_2(\RR^3,\CC^{2s+1},d\nu)$.
In such a representation the generators of $U\mid_{{\mathcal P}_+^\uparrow}$ are in
the so called {\sl canonical form}
$$
P_j=\left[\begin{array}{cc}p_j&0\cr 0&p_j\end{array}\right],\quad
P_0=\left[\begin{array}{cc}p_0&0\cr 0&-p_0\end{array}\right],\quad
J_k=\left[\begin{array}{cc}{\textsf j}_k&0\cr 0&{\textsf j}_k\end{array}\right],\quad
K_j=\left[\begin{array}{cc}{\textsf k}_j&0\cr 0&-{\textsf k}_j\end{array}\right],\eqno(2)
$$
where
${\textsf j}_k=i\left(p_l\frac{\partial}{\partial p_j}-p_j\frac{\partial}{\partial p_l}\right)+S_k$ and
${\textsf k}_j=ip_0\frac{\partial}{\partial p_j}-\frac{({\bf S}\land {\bf P})_j}{\mu+p_0}$.
\vskip.5pc\noindent
Then there are the following six inequivalent representations $U^{(1)}$, $U^{(2)}$,...,$U^{(6)}$ with the same $s$, $\mu$.
\begin{itemize}
\item[] $U^{(1)}$ has unitary $\TREV=\left[\begin{array}{cc}0&1\cr 1&0\end{array}\right]$ and unitary $\SREV=\Upsilon\left[\begin{array}{cc}1&0\cr 0&1\end{array}\right]$;
\item[] $U^{(2)}$ has unitary $\TREV=\left[\begin{array}{cc}0&1\cr 1&0\end{array}\right]$ and unitary $\SREV=\Upsilon\left[\begin{array}{cc}1&0\cr 0&-1\end{array}\right]$;
\item[] $U^{(3)}$ has unitary $\TREV=\left[\begin{array}{cc}0&1\cr 1&0\end{array}\right]$ and
anti-unitary $\SREV=\left[\begin{array}{cc}0&\tau\cr \tau&0\end{array}\right]{\mathcal K}$;
\item[] $U^{(4)}$ has unitary $\TREV=\left[\begin{array}{cc}0&1\cr 1&0\end{array}\right]$ and
anti-unitary $\SREV=\left[\begin{array}{cc}0&\tau\cr -\tau&0\end{array}\right]{\mathcal K}$;
\item[] $U^{(5)}$ has anti-unitary $\TREV=\tau{\mathcal K}\Upsilon\left[\begin{array}{cc}0&1\cr 1&0\end{array}\right]$ and anti-unitary$\SREV=\left[\begin{array}{cc}0&\tau\cr\tau&0\end{array}\right]{\mathcal K}$;
\item[] $U^{(6)}$ has anti-unitary $\TREV=\tau{\mathcal K}\Upsilon\left[\begin{array}{cc}0&1\cr 1&0\end{array}\right]$ and anti-unitary$\SREV=\left[\begin{array}{cc}0&\tau\cr-\tau&0\end{array}\right]{\mathcal K}$.
\end{itemize}
(the matrix entries ``$1$'' and ``$0$'' denote the identity and null operators of  $\CC^{2s+1}$).
\vskip.5pc\noindent
Thus, fixed $s$ and $\mu$,
taking into account
all cases ({\bf u}), ({\bf d}), ({\bf s}), there are
eight inequivalent irreducible generalized projective representations
of $\mathcal P$ with $U^\pm\mid_{{\mathcal P}_+^\uparrow}$ irreducible.
However, all these octets do not exhaust the class ${\mathcal I}_{\mathcal P}$ of all irreducible generalized projective representations of $\mathcal P$ with these $s$, $\mu$, because the class of irreducible representations of $\mathcal P$ with $U^+\mid_{{\mathcal P}_+^\uparrow}$ or  $U^-\mid_{{\mathcal P}_+^\uparrow}$ reducible is not empty, as shown in \cite{PROOFS}; in fact,
the whole class ${\mathcal I}_{\mathcal P}$ contains classes that are not considered in the literature about relativistic quantum theories of single particles,
namely
\vskip.5pc\noindent
${\mathcal I}_{\mathcal P}({\rm ant.})$, i.e. the class that collects all representation of the kind $U^{(3)}$-$U^{(6)}$,
\vskip.5pc\noindent
${\mathcal I}_{\mathcal P}(U^\pm{\rm red.})$, i.e. the class of all representations in ${\mathcal I}_{\mathcal P}$ with $U^+\mid_{{\mathcal P}_+^\uparrow}$ or  $U^-\mid_{{\mathcal P}_+^\uparrow}$ reducible.
\section{Quantum theories of single particles}
In order to identify the specific theories of isolated systems, we interpret $\mathcal P$ as a group of changes of reference frame, according to special relativity.
Accordingly,
given a reference frame $\Sigma$ in the class $\mathcal F$ of the (inertial) reference frames that move uniformly with respect to each other,
for every $g\in{\mathcal P}$, $\Sigma_g$ denotes the reference frame
related to $\Sigma$ just by $g$, and let $\textsf g:\RR^4\to\RR^4$ be the mapping such that if
$\underline x=(t,x_1,x_2,x_3)\equiv(x_0,{\bf x})$ is the vector of the time-space coordinates of an event with respect
to $\Sigma$, then $\textsf g(\underline x)$ is the vector of the time-space coordinates of that event with respect to $\Sigma_g$.
\par
Let us now consider an {\sl isolated} physical system. We formulate the following statement as a {\sl physical principle} valid for this system.
\begin{description}
\item[($\mathcal S${\it ym})]{\sl
The theory of an isolated system is invariant for changes of frames in $\mathcal F$.}
\end{description}
Now we imply from ($\mathcal S${\it ym}) that for each transformation $g\in\mathcal P$ a specific {\sl quantum transformation}
$S_g:\Omega({\mathcal H})\to\Omega({\mathcal H})$, $A\to S_g[A]$ of the quantum observables exists,
we shall define
below through the concept ($\mathcal I${\it nd}) of
{\sl relative indistinguishability} between measuring procedures of quantum observables.
\begin{itemize}
\item[($\mathcal I${\it nd})]
{\sl Given two reference frames $\Sigma_1$ and $\Sigma_2$ in $\mathcal F$, if a measuring procedure ${\mathcal M}_1$ is relatively to $\Sigma_1$ identical to what is another measuring procedure
${\mathcal M}_2$ relatively to $\Sigma_2$, we say that ${\mathcal M}_1$ and ${\mathcal M}_2$ are indistinguishable
relatively to $(\Sigma_1,\Sigma_2)$.}
\end{itemize}
Given $\Sigma_1$ and $\Sigma_2$ in $\mathcal F$,
for every measuring procedure ${\mathcal M}_1$ another measuring procedure ${\mathcal M}_2$ must exist such that
${\mathcal M}_1$ and ${\mathcal M}_2$ are indistinguishable
relatively to $(\Sigma_1,\Sigma_2)$, otherwise the invariance stated by ($\mathcal S${\it ym}) would fail.
\vskip.5pc\noindent
{\bf Definition 4.1.} ($\mathcal{QT}$).
{\sl Fixed any $\Sigma\in{\mathcal F}$,
the {\it quantum transformation} of
$g\in{\mathcal P}$ is the mapping
$$
S_g:\Omega({\mathcal H})\to \Omega({\mathcal H}),\quad A\to S_g[A]\,,\eqno(3)
$$
such that
the quantum observables $A$ and $S_g[A]$ are measured by two measuring procedures
${\mathcal M}_1$ and ${\mathcal M}_2$, respectively, that are indistinguishable relatively to $(\Sigma,\Sigma_g)$.}
\vskip.5pc\noindent
Then, the physical principles compel \cite{PROOFS} the following properties of quantum transformations.
\begin{itemize}
\item[(S.1)] {\sl Every $S_g:\Omega({\mathcal H})\to\Omega({\mathcal H})$ is bijective.}
\item[(S.2)] {\sl For every function $f$ the equality $S_g[f(A)]=f(S_g[A])$ holds for all $A\in\Omega({\mathcal H})$.}
\item[]
Indeed,
in general the quantum observable $D=f(C)$ can be measured by transforming the outcome $c$ of the
procedure for $C$ into the outcome $f(c)=d$ of $D$; now, the application of the same $f$ to the outcomes of ${\mathcal M}_1$ and ${\mathcal M}_2$,
relatively indistinguishable procedures for $A$ and $S_g[A]$, yield
two procedures ${\mathcal N}_1$ and ${\mathcal N}_2$ that preserve relative indistinguishability;
so, ${\mathcal N}_2$ measures $f(S_g[A])$, but also $S_g[f(A)]$ being ${\mathcal N}_1$ and ${\mathcal N}_2$ relatively indistinguishable.
\item[(S.3)] {\sl $S_{gh}=S_g\circ S_h$, for all $g,h\in\mathcal P$.}
\end{itemize}
Properties (S.1) and (S.2) were sufficient  \cite{N1} to prove that each quantum transformation $S_g$ is an {\sl automorphism} of the lattice $\Pi({\mathcal H})$
of the projection operators; therefore, according to Wigner theorem \cite{N1},\cite{b2}, a unitary or anti-unitary operator
${\tilde U}_{g}$ must exist such that
$$
S_{g}[A]={\tilde U}_gA{\tilde U}_g^{-1}\,,\quad\hbox{for every }A\in\Omega({\mathcal H}).
\eqno(4)
$$
Given any real function $\theta$ of $g$, the operators $\tilde U_g$ and $e^{i\theta(g)}\tilde U_g$
yield the same quantum transformation as $\tilde U_g$, i.e.
$\tilde U_gA \tilde U_g^{-1}= \left(e^{i\theta(g)}\tilde U_g\right)A\left(e^{i\theta(g)}\tilde U_g\right)^{-1}$,
and, hence, $U_g =e^{i\theta(g)}\tilde U_g$ can replace $\tilde U_g$ in the specific quantum theory of the system.
In particular, we can set $U_e=\Id$.
\vskip.5pc
Condition (S.3) implies that $U_{g_1g_2}=\varsigma(g_1,g_2)U_{g_1}U_{g_2}$.
Hence, in general the correspondence $U:{\mathcal P}\to {\mathcal V}({\mathcal H})$, $g\to U_g$ realized according to these prescriptions is a
{\sl generalized} projective representation. We assume
that the correspondence $g\to S_g$ from ${\mathcal P}_+^\uparrow$ into the automorphisms of $\Pi({\mathcal H})$
is continuous, according to Bargmann topology \cite{N1}; this assumption can be motivated by the idea that a small transformation $g\in{\mathcal P}_+^\uparrow$
determine a small change from $A$ to $S_g[A]$. This continuity implies \cite{PROOFS} that
the restriction $U:{\mathcal P}_+^\uparrow\to {\mathcal U}({\mathcal H})$ can be made continuous by redefining the free phase $\theta(g)$,
and therefore it is a continuous projective representation, according to Prop. 2.1.
\vskip.5pc
Thus, from the principles ($\mathcal S${\it ym}) we have inferred
that the Hilbert space of the quantum theory of an isolated system must necessarily be the Hilbert space of a generalized projective representation
of $\mathcal P$, that determines the quantum transformations as $S_g[A]=U_gAU_g^{-1}$; moreover, the restriction $U\mid_{{\mathcal P}_+^\uparrow}$
is continuous.
\subsection{Localizable particle theories}
By {\sl localizable free particle}, shortly {\sl free particle}, we
mean an isolated system whose quantum theory is endowed with a unique {\sl position observable}, namely with a {\sl unique} triple
$(Q_1,Q_2,Q_3)\equiv{\bf Q}$ of self-adjoint operators,
whose components $Q_j$ are called coordinates, characterized by the following conditions.
\begin{itemize}
\item[({\it Q}.1)]
$[Q_j,Q_k]=\nop$, for all $j,k=1,2,3$.
\item[]
This condition establishes that a measurement on a particle is possible, that yields
all three values of the coordinates of that specimen of the particle.
\item[({\it Q}.2)]
The triple $(Q_1,Q_2,Q_3)\equiv{\bf Q}$ is characterized by
the specific properties of transformation of position with respect to the group $\mathcal P$,
i.e. by the specific mathematical relations between $\bf Q$ and $S_g[{\bf Q}]$.
\end{itemize}
{\bf Example 4.1.} Condition ({\it Q}.2) implies that the following statements hold.
\begin{itemize}
\item[({\it Q}.2]\hskip-3pt.a)\quad
$S_\trev[{\bf Q}]={\bf Q}$ and $S_\srev[{\bf Q}]=-{\bf Q}$, equivalent to $\TREV{\bf Q}={\bf Q}\TREV$ and $\SREV{\bf Q}=-{\bf Q}\SREV$.
\item[({\it Q}.2]\hskip-3pt.b)\quad
If $g\in\mathcal E$ then $S_g[{\bf Q}]=U_g{\bf Q}U_g^{-1}={\textsf g}({\bf Q})$, ${\bf x}\to{\textsf g}({\bf x})$ being the function that realizes $g$.
\end{itemize}
Hence, if $g$ is a
translation along $x_1$, then $S_g[Q_k]=e^{-iP_1a}Q_ke^{iP_1a}=Q_k-\delta_{1k}a$, which implies
$$
[Q_k,P_j]=i\delta_{jk}\;.\eqno(5.i)
$$
Analogously, the transformation properties with respect to spatial rotations imply
$$
[J_l,Q_j]=i\hat\epsilon_{ljk}Q_k, \eqno(5.ii)
$$
where $\hat\epsilon_{ljk}$ is the Levi-Civita symbol restricted by the condition $l\neq j\neq k$.
\vskip.5pc
Following a customary habit, we refer to a particle for which $U$
is {\sl irreducible} as an {\sl elementary particle} \cite{PROOFS}.
Accordingly,
by selecting the irreducible generalized projective representations $U$
of $\mathcal P$ that admit such a triple $\bf Q$ we identify the possible theories of elementary free particles.
\par
In this section we identify
complete theories for the case with spin parameter $s=0$.
The case $s>0$ is addressed in \cite{PROOFS}.
\subsection{Elementary particle: cases $s=0$, $\sigma(P_0)=I_\mu^\pm$ and $U^\pm\mid_{{\mathcal P}_+^\uparrow}$ irreducible}
Let $U$ be any irreducible representation of $\mathcal P$ with $\sigma(P_0)=I_\mu^+$, given in section 3.1, where ${\mathcal H}=L_2(\RR^3,\CC^{2s+1}d\nu)$.
The Newton and Wigner self-adjoint operators \cite{N-W} are
$F_j=i\frac{\partial}{\partial p_j}-\frac{i}{2p_0^2}p_j$.
Since $[F_j,P_k]=i\delta_{jk}$ holds, by (5.i) we imply
$Q_j-F_j=f_j(\bf P)$, i.e. $\left((Q_j-F_j)\psi\right)({\bf p})=f_j({\bf p})\psi({\bf p})$, where $f_j({\bf p})\in\Omega(\CC^{2s+1})$ for all ${\bf p}\in\RR^3$.
On the other hand, since $[J_j,F_k]=i{\hat\epsilon}_{jkl}F_l$, by (5.ii) we imply
$$
[J_j,f_k({\bf P})]=i{\hat\epsilon}_{jkl}f_l({\bf P}).\eqno(6)
$$
In case $s=0$, we have $S_k=0$, $\Omega(\CC^{2s+1})=\Omega(\CC)\equiv\RR$,  and $\tau=1$. Then (6)
together with $[J_j,P_k]=i{\hat\epsilon}_{jkl}P_l$ implies $f_j({\bf P})=h(\vert{\bf p}\vert)p_j$; by redefining $h(\vert{\bf p}\vert)=f(p_0)$, with
$p_0=\sqrt{\mu^2+{\bf p}^2}$, we have
$$
{\bf f}({\bf P})=f(p_0){\bf p},\hbox{ where }f(p_0)\in\Omega(\CC)\equiv\RR.\eqno(7)
$$
Now,  $\TREV F_j=F_j\TREV$ straightforwardly holds and ({\it Q}.2.a) implies $\TREV Q_j=Q_j\TREV$, so that we obtain
$\TREV f_j({\bf p})=f_j({\bf p})\TREV$; from this equality, since $\TREV={\mathcal K}\Upsilon$, by (7) we derive
$$
\overline{f(p_0)}=-f(p_0).\eqno(8)
$$
Since $f(p_0)\in\RR$, (8) implies $f(p_0)=0$.
Therefore, ${\bf Q}={\bf F}$. The condition $S_\srev[{\bf Q}]=-{\bf Q}$, i.e. $\SREV{\bf Q}=-{\bf Q}\SREV$ turns out to be trivially satisfied.
\par
Thus, if $s=0$, there is a unique position operator $\bf Q={\bf F}$
and it is completely determined by (Q.1) and ({\it Q}.2.a),({\it Q}.2.b).
This result agrees with the different derivations of the Newton and Wigner operators as position operators \cite{N-W},\cite{J80}.
\vskip.5pc\noindent
By a quite similar derivation the same result, ${\bf Q}={\bf F}$ is obtained for $\sigma(P_0)=I_\mu^-$
\subsection{The case $s=0$, $\sigma(P_0)=I_\mu^+\cup I_\mu^-$ and $U^\pm\mid_{{\mathcal P}_+^\uparrow}$ irreducible}
Let $U$ be any irreducible representation of $\mathcal P$ with $\sigma(P_0)=I_\mu^+\cup I_\mu^+$ and $s=0$ of section 3.2.
Let us define
${\bf D}={\bf Q}-\hat{\bf F}$, where
$\hat{\bf F}={\bf F}\Id\equiv\left[\begin{array}{cc}{\bf F}&0\cr 0&{\bf F}\end{array}\right]$ is the Newton-Wigner operator in this representation.
Conditions ({\it Q}.2.b) for $g\in{\mathcal E}$ imply \cite{PROOFS}
$$
Q_j=F_j+D_j,\hbox{ where }D_j=\left[\begin{array}{cc}d_{11}(p_0)&d_{12}(p_0)\cr d_{21}(p_0)&d_{22}(p_0)\end{array}\right]p_j\;\hbox{ and }\; d^\ast_{mn}(p_0)=d_{nm}(p_0).\eqno(9)
$$
So, to determine $\bf Q$ we have to determine the functions $d_{mn}$ of $p_0$; the conditions $\TREV{\bf Q}={\bf Q}\TREV$ and $\SREV{\bf Q}=-{\bf Q}\SREV$
can help in solving the indeterminacy. However, according to section 3.2,
now the explicit form of $\TREV$ and $\SREV$ depend on their unitary or anti-unitary character; so we show the result determined in \cite{PROOFS}
for each different combination \cite{PROOFS}.
\vskip.8pc\noindent
{\bf (UU)} Let us start with the case where
$\TREV=\left[\begin{array}{cc}0&1\cr 1&0\end{array}\right]$, while $\SREV=\Upsilon\left[\begin{array}{cc}1&0\cr 0&1\end{array}\right]$
or $\SREV=\Upsilon\left[\begin{array}{cc}1&0\cr 0&-1\end{array}\right]$. By making use of these explicit forms and of ({\it Q}.2a) we found that
$\bf D$ and hence ${\bf Q}$ remain undetermined.
\vskip.8pc\noindent
{\bf (UA)} In the case that
$\TREV=\left[\begin{array}{cc}0&1\cr 1&0\end{array}\right]$, while $\SREV={\mathcal K}\left[\begin{array}{cc}0&1\cr 1&0\end{array}\right]$
or $\SREV={\mathcal K}\left[\begin{array}{cc}0&1\cr -1&0\end{array}\right]$, we found that
\vskip.5pc\noindent
i) if $\SREV={\mathcal K}\left[\begin{array}{cc}0&1\cr -1&0\end{array}\right]$, then $\bf Q$ is still undetermined.
\vskip.5pc\noindent
ii) if $\SREV={\mathcal K}\left[\begin{array}{cc}0&1\cr 1&0\end{array}\right]$, then $\bf Q$ is uniquely determined, and ${\bf Q}=\hat{\bf F}$.
\vskip.8pc\noindent
{\bf (AA)} In the case
$\TREV={\mathcal K}\Upsilon\left[\begin{array}{cc}1&0\cr 0&1\end{array}\right]$, while $\SREV={\mathcal K}\left[\begin{array}{cc}0&1\cr 1&0\end{array}\right]$
or $\SREV={\mathcal K}\left[\begin{array}{cc}0&1\cr -1&0\end{array}\right]$,
we found that\par\noindent
i) if $\SREV={\mathcal K}\left[\begin{array}{cc}0&1\cr -1&0\end{array}\right]$ then $\bf Q$ is still undetermined.
\vskip.5pc\noindent
ii) if $\SREV={\mathcal K}\left[\begin{array}{cc}0&1\cr 1&0\end{array}\right]$ then $\bf Q$ is uniquely determined and ${\bf Q}=\hat{\bf F}$.
\subsection{Klein-Gordon particles}
In sections 4.2 and 4.3, in the case $s=0$,
for every value of the characterizing parameter $\mu>0$ four inequivalent theories of single particle have been singled out with $\bf Q$ determined by
({\it Q}.2.a) and ({\it Q}.2b).
To complete the theories, we determine the explicit form of the wave equations, by replacing $P_0$ with the specific time translation operator,
explicitly known in each specific theory.
\par
The so completed theories can be re-formulated in the following equivalent forms, obtained by means of unitary transformations
operated by the unitary operator $Z=Z_1Z_2$, where $Z_2=\frac{1}{\sqrt{p_0}}\Id$ and $Z_1$ is the inverse of the {\sl Fourier-Plancherel} operator, that transforms $\psi({\bf p})$ into $(Z\psi)({\bf x})\equiv(\hat\psi)({\bf x})$.
\begin{itemize}
\item[$\mathcal T$.1]
The theory of section 4.2, identified by $\sigma(P_0)=I_\mu^+$ and $s=0$,
can be equivalently reformulated in the Hilbert space
${\mathcal H}=Z\left(L_2(\RR^3,d\nu)\right)\equiv L_2(\RR^3)$.
Here the self-adjoint generators are ${\hat P}_j=ZP_jZ^{-1}=-i\frac{\partial}{\partial x_j}$, ${\hat P}_0=\sqrt{\mu^2+{\nabla}^2}$,
${\hat J}_k=-i\left(x_l\frac{\partial}{\partial x_j}-x_j\frac{\partial}{\partial x_l}\right)$,
${\hat K}_j=\frac{1}{2}(x_j{\hat P}_0+{\hat P}_0x_j)$,
while ${\hat \SREV}=\Upsilon$, ${\hat \TREV}={\mathcal K}$. The Newton-Wigner operator representing position, in this representation becomes
the multiplication operator ${\hat Q}_j$, defined by ${\hat Q}_j\psi({\bf x})=x_j\psi({\bf x})$.
Accordingly, the wave equation is
$$
i\frac{\partial}{\partial t}\psi_t({\bf x})=\sqrt{\mu^2-\nabla^2}\psi_t({\bf x})\,.\eqno{(10)}
$$
\item[$\mathcal T$.2]
The new formulation of
the theory of section 4.2, identified by $\sigma(P_0)=I_\mu^-$ and $s=0$,
differs from $\mathcal T$.1 just for $P_0$ and $K_j$ that change sign, so that
$$
i\frac{\partial}{\partial t}\psi_t({\bf x})=-\sqrt{\mu^2-\nabla^2}\psi_t({\bf x}).\eqno{(11)}
$$
\item[$\mathcal T$.3]
The theory of section 4.3, with $\sigma(P_0)=I_\mu^+\cup I_\mu^-$ and $s=0$,
identified by $\TREV=\left[\begin{array}{cc}0&1\cr 1&0\end{array}\right]$
and
$\SREV={\mathcal K}\left[\begin{array}{cc}0&1\cr 1&0\end{array}\right]$, can be equivalently reformulated in the Hilbert space
${\mathcal H}=Z\left(L_2(\RR^3,d\nu)\oplus L_2(\RR^3,d\nu)\right)\equiv L_2(\RR^3)\oplus L_2(\RR^3)$; the new self-adjoint generators are
${\hat P}_j=\left[\begin{array}{cc}-i\frac{\partial}{\partial x_j}&0\cr 0&-i\frac{\partial}{\partial x_j}\end{array}\right]$,
${\hat P}_0=\sqrt{\mu^2-\nabla^2}\left[\begin{array}{cc}1&0\cr 0&-1\end{array}\right]$,
${\hat J}_k=-i\left(x_l\frac{\partial}{\partial x_j}-x_j\frac{\partial}{\partial x_l}\right)\left[\begin{array}{cc}1&0\cr 0& 1\end{array}\right]$;
${\hat K}_j=\frac{1}{2}\left(x_j\sqrt{\mu^2-\nabla^2}+\sqrt{\mu^2-\nabla^2}x_j\right)\left[\begin{array}{cc}1&0\cr 0&-1\end{array}\right]$.
The position operator is
${\hat Q}_j=\left[\begin{array}{cc}x_j&0\cr 0&x_j\end{array}\right]$.
\par\noindent
The wave equation is
$$
i\frac{\partial}{\partial t}\left[\begin{array}{c}\psi^+_t({\bf x})\cr \psi^-_t({\bf x})\end{array}\right]
=\left[\begin{array}{c}\sqrt{\mu^2-\nabla^2}\psi^+_t({\bf x})\cr -\sqrt{\mu^2-\nabla^2}\psi^-_t({\bf x})\end{array}\right]\,.\eqno(12)
$$
\item[$\mathcal T$.4]
The theory corresponding to ({\bf AA}.ii) in section 4.3,
identified by $\TREV=\Upsilon{\mathcal K}\left[\begin{array}{cc}1&0\cr 0&1\end{array}\right]$
and
$\SREV={\mathcal K}\left[\begin{array}{cc}0&1\cr 1&0\end{array}\right]$,
differs from $\mathcal T$.3 only for these operators.
\end{itemize}
\vskip.5pc\noindent
The early theory \cite{Klein},\cite{Fock},\cite{Gordon}, for spin 0 particle establishes that the wave equation is Klein-Gordon equation
$$
\left(\frac{\partial^2}{\partial t^2}-\nabla^2\right)\psi_t({\bf x})=
-m^2\psi_t({\bf x})\,,\eqno{(13)}
$$
which is second order with respect to time.
This is an evident difference with respect to theories $\mathcal T$.1-$\mathcal T$.4,
where all wave equations are first order.
However, in each theory $\mathcal T$.1-$\mathcal T$.4 if the wave equation is solved by $\psi_t$, then the derivative of the
equation with respect to time yields
$-\frac{\partial^2}{\partial t^2}\psi_t=iP_0\frac{\partial}{\partial t}\psi_t=P_0^2\psi_t$,
since $\frac{\partial}{\partial t}$ commutes with $P_0$ in all cases, obtaining
$$
\frac{\partial^2}{\partial t^2}\psi_t({\bf x})-{\nabla}^2\psi_t({\bf x})=-\mu^2\psi_t({\bf x})\,,\eqno(14)
$$
which coincides with
Klein-Gordon equation,
once identified $\mu$ with the mass $m$. However, this coincidence does not mean that theories $\mathcal T$.1-$\mathcal T$.4
are equivalent to Klein-Gordon theory.
A first difference is that according to our approach there are {\sl four} inequivalent theories for spin 0 and ``mass'' $\mu$ particles.
In $\mathcal T$.1 there are no wave functions corresponding to negative spectral values of $P_0$. In $\mathcal T$.2 the positive values are
forbidden. Klein-Gordon theory does not exhibit this differentiation. In particular, the space of the vector states is only one,
namely the space generated by the solutions of (13).
\par
A second evidence of non-equivalence is the difference between the set of solutions of the
respective wave equations:
while all solutions of the wave equations of $\mathcal T$.1-$\mathcal T$.4 are solution of Klein-Gordon equation,
the converse is not true, in general; thus, from the point of view of $\mathcal T$.1-$\mathcal T$.4,
the extra solutions of Klein-Gordon equation are {\sl un-physical}.
\vskip.5pc
A third important difference concerns with the physical interpretation and its consistency.
By means of mathematical manipulation it can be implied that for every solution $\psi_t$ of Klein-Gordon equation (13)
the ``continuity'' equation $\frac{\partial}{\partial t}\hat \rho ={\mathbf\nabla}\cdot{\hat{\bf j}}$ holds,
where
$\hat\rho(t,{\bf x})=\frac{i}{2m}\left(\overline{\psi_t}\frac{\partial}{\partial t}\psi_t-{\psi_t}\frac{\partial}{\partial t}\overline{\psi_t}\right)$
was interpreted as the {\sl probability of position} density and
$\hat{\bf j}(t,{\bf x})=\frac{i}{2m}\left(\psi_t\nabla\overline{\psi_t}-\overline{\psi_t}\nabla\psi_t\right)$ as
its {\sl current} density.
This interpretation is at the basis of the Dirac concern \cite{mehra} that position probability density can be negative.
A way to overcome the difficulty was proposed by Feshbach and Villars \cite{FV}.
They derive an equivalent form of Klein-Gordon equation as a first order equation
$i\frac{\partial}{\partial t}\Psi_t=H\Psi_t$
for the state vector $\Psi_t=\left[\begin{array}{c}\phi_t\cr \chi_t\end{array}\right]$, where
$\phi_t=\frac{1}{\sqrt{2}}(\psi_t+\frac{1}{m}\frac{\partial}{\partial t}\psi_t)$,
$\chi_t=\frac{1}{\sqrt{2}}(\psi_t-\frac{1}{m}\frac{\partial}{\partial t}\psi_t)$, and
$H=(\sigma_3+\sigma_2)\frac{1}{2m}(\nabla+m\sigma_3)$;
in this representation $\hat\rho=\vert\phi_t\vert^2-\vert\chi_t\vert^2$, without time derivatives.
The minus sign in $\hat\rho$ forbids to interpret it as probability density of position;
Feshbach and Villars proposed to interpret it as {\sl density probability of charge}.
Nevertheless, according to Barut and Malin \cite{BM68}, covariance with respect to boosts should imply that
$\hat\rho$ must be the time component of a four-vector.
Barut and Malin proved that is not the case.
\vskip.5pc
The theories $\mathcal T$.1-$\mathcal T$.4 do not suffer these problems. Indeed, in all of them the position is represented by the multiplication operator;
therefore,
the probability density of position must necessarily be $\rho(t,{\bf x})=\vert\psi_t({\bf x})\vert^2$ in $\mathcal T$.1, $\mathcal T$.2
and $\rho(t,{\bf x})=\vert\psi_t^+({\bf x})\vert^2+\vert\psi_t^-({\bf x})\vert^2$ in $\mathcal T$.3, $\mathcal T$.4.
Thus, the quantum state at a given time determines the {\sl non-negative} probability density of position.
\par
On the other hand, the covariance properties with respect to boosts, according to ({\it Q}.2), are explicitly expressed by
$S_g[{\bf Q}]=e^{iK_j\varphi(u)}{\bf Q}e^{-iK_j\varphi(u)}$, being $K_j$ and $\bf Q$ explicitly known,
and there is no need of a four-density concept.
\par
Each of the theories $\mathcal T$.1-$\mathcal T$.4 corresponds to a possible
kind of particle; being unitarily inequivalent, they correspond to different kind of particles. The actual existence in nature
of each of these particles is not a matter that can be assessed in this theoretical paper.

\end{document}